\documentclass[]{iopart}

\usepackage{graphicx}

\begin{document}

\title{Multipolar analysis of spinning binaries}

\author{E. Berti$^{1,2}$, V. Cardoso$^{3,4}$, J. A. Gonz{\'a}lez$^{5,6}$,
U. Sperhake$^5$, B. Br{\"u}gmann$^5$}

\address{$^1$ Jet Propulsion Laboratory, California Institute of Technology,
         Pasadena, CA 91109, USA}

\address{$^2$ McDonnell Center for the Space Sciences, Department of Physics,
         Washington University, St.~Louis, MO 63130, USA}

\address{$^3$~Department of Physics and Astronomy,
         The University of Mississippi, University, MS 38677-1848, USA}

\address{$^4$~Centro Multidisciplinar de Astrof\'{\i}sica - CENTRA,
  Departamento de F\'{\i}sica, Instituto Superior T\'ecnico, Av. Rovisco Pais
  1, 1049-001 Lisboa, Portugal}

\address{$^5$~Theoretisch Physikalisches Institut,
         Friedrich Schiller Universit\"at, 07743 Jena, Germany}

\address{$^6$~Instituto de F\a'{\i}sica y Matem\a'aticas,
         Universidad Michoacana de San Nicol\a'as de Hidalgo,
         C. P. 58040 Morelia, Michoac\a'an, M\a'exico}

\ead{berti@wugrav.wustl.edu}

\begin{abstract}
  We present a preliminary study of the multipolar structure of gravitational
  radiation from spinning black hole binary mergers. We consider three
  different spinning binary configurations: (1) one ``hang-up'' run, where the
  black holes have equal masses and large spins initially aligned with the
  orbital angular momentum; (2) seven ``spin-flip'' runs, where the holes have
  a mass ratio $q\equiv M_1/M_2=4$, the spins are anti-aligned with the orbital
  angular momentum, and the initial Kerr parameters of the holes $j_1=j_2=j_i$
  (where $j\equiv J/M^2$) are fine-tuned to produce a Schwarzschild remnant
  after merger; (3) three ``super-kick'' runs where the mass ratio $q=1,~2,~4$
  and the spins of the two holes are initially located on the orbital plane,
  pointing in opposite directions.  For all of these simulations we compute
  the multipolar energy distribution and the Kerr parameter of the final hole.
  For the hang-up run, we show that including leading-order spin-orbit and
  spin-spin terms in a multipolar decomposition of the post-Newtonian
  waveforms improves agreement with the numerical simulation.
\end{abstract}



These are exciting times for gravitational wave (GW)
research. Earth-based laser-interferometric detectors are collecting
data at design sensitivity, and LIGO \cite{Abramovici:1992ah} just
completed the longest scientific run to date. The space-based
interferometer LISA is expected to open an observational window at low
frequencies ($\sim 10^{-4}-10^{-1}$~Hz) within the next decade
\cite{Danzmann:1998}.  Following remarkable breakthroughs in the
simulation of the strongest expected GW sources, the inspiral and
coalescence of black hole binaries
\cite{Pretorius2005a,Campanelli2006, Baker2006}, several groups have
now explored various aspects of this problem, including
spin-precession and spin-flips
\cite{Campanelli2007b,Campanelli2007v2}, comparisons of numerical
results with post-Newtonian (PN) predictions
\cite{Buonanno2006, Berti:2007fi, Hannam2007, Boyle:2007ft},
multipolar analyses of
the emitted radiation \cite{Buonanno2006, Berti:2007fi,
  Schnittman:2007ij} and the use of numerical waveforms in data
analysis \cite{Baumgarte:2006en,Pan2007,Ajith:2007qp,Ajith:2007kx}.

In Ref.~\cite{Berti:2007fi} we studied the multipolar distribution of
radiation and the final spin resulting from the merger of
unequal-mass, non-spinning black holes with mass ratios $q=M_1/M_2$ in
the range 1 to 4.  The main purpose of this paper is to show
preliminary results from our attempt to extend the analysis to
spinning binaries.

A second purpose of this study is to test recent predictions for the
spin of the black hole resulting from a generic merger.  Buonanno,
Kidder and Lehner \cite{Buonanno:2007sv} recently introduced a
surprisingly accurate model, based on the extrapolation of
point-particle results, that was shown to be in good agreement with
existing numerical simulations (see also \cite{Rezzolla:2007rd}). An
interesting question explored in \cite{Buonanno:2007sv} concerns
spin-flip configurations.  Suppose that initially both black holes
have equal Kerr parameters ($j_i=j_1=j_2$), and spins {\em
  antialigned} with respect to the orbital angular momentum. For a
given mass ratio $q$, which value of $j_i$ will produce a
Schwarzschild black hole? These ``critical'' configurations could be
very interesting, since mild variations of the parameters around the
critical values may produce interesting orbital dynamics (eg.~spin
flips) and complex gravitational waveforms. As argued in
\cite{Buonanno:2007sv}, one needs unequal masses to be able to produce
a Schwarzschild remnant at all. For $q=4$ and zero eccentricity,
Ref.~\cite{Buonanno:2007sv} predicts that a Schwarzschild black hole
should be formed when $j_i=-0.815$. A semi-analytical fitting formula
\cite{Rezzolla:2007rd} predicts a critical spin $j_i=-0.823$.  Here we
provide a numerical benchmark against which to test these analytical
models, and possibly other models that may be developed in the future,
by computing the final Kerr parameter $j_{\rm fin}$ from a sequence of
spinning binaries with $q=4$ and values of $j_i\in [-0.75,-0.87]$. The
numerical results are well fitted by a linear relation of the form
$j_{\rm fin} = -0.570(j_i-0.842)$.  From this fit, our best
estimate for the initial spin leading to the formation of a
Schwarzschild remnant is $j_i\simeq -0.842\pm0.003$.

The plan of the paper is as follows. In Sec.~\ref{simulations} we
introduce the numerical code and we list the new simulations
considered in this paper.  In Sec.~\ref{nospin} we compare the
multipolar energy distribution of spinning and non-spinning binaries.
In Sec.~\ref{spin} we generalize our multipolar decomposition of PN
waveforms to include leading-order spin contributions in the special
case of spins aligned (or anti-aligned) with the orbital angular
momentum, and we show (in a special case) that the inclusion of spin
terms improves the agreement with numerical results
\cite{Marronetti:2007wz}.  In Sec.~\ref{Schw} we study in detail
``spin-flip'' configurations designed to produce a Schwarzschild
remnant.  Finally, in Sec.~\ref{EMOP} we discuss the fraction of
energy radiated in ringdown in the different simulations.


\section{Numerical simulations}
\label{simulations}

In this work we compare two sequences of numerical black hole binary
simulations.

Sequence 1 is a series of simulations of non-spinning black hole
binaries with mass ratio $q$ ranging from $1$ to $4$. These
simulations were performed with the {\sc Bam} code
\cite{Bruegmann2006a}, and they were used in \cite{Gonzalez2007} to
study gravitational recoil and in \cite{Berti:2007fi} to investigate
the multipolar structure of the emitted gravitational radiation.
Sequence 2 consists of simulations of binary systems with mass-ratios
in the same range, but with non-vanishing spins. In particular, we
study two families of spinning binaries.  For the first family, the
spin of both holes is either aligned (run uu1) or anti-aligned (runs
dd1--dd7) with the orbital angular momentum. For the second
family, the spins of the two holes are initially located on the
orbital plane, and they point in opposite directions.  The latter
configuration produces surprisingly large recoil velocities
\cite{Campanelli2007v2, Gonzalez2007a}, and it is sometimes referred
to as a ``super-kick'' configuration (runs sk1, sk2, sk4).

Sequence 2 binaries have been evolved with an advanced version of the
{\sc Lean} code described in \cite{Sperhake2006}. The key improvement
over the original code is the implementation of sixth-order accurate
stencils for spatial derivatives, as introduced in
\cite{Husa2007b}. {\sc Lean} is based on the {\sc Cactus} computational
toolkit and uses {\sc Carpet} \cite{Schnetter2004} for mesh-refinement,
{\sc TwoPunctures} \cite{Ansorg2004} for puncture initial data
and {\sc AHFinderDirect} \cite{Thornburg1996, Thornburg2004} for
horizon finding.
We have used the advanced version of {\sc Lean} for
all spinning configurations, except for the model labelled uu1 in
Table \ref{tab: models}, which has first been reported in
Ref.~\cite{Marronetti:2007wz}.

\begin{table}[t]
  \centering \caption{\label{tab: models} Details of the spinning binary
    simulations of sequence 2. The holes start on the $x$ axis with a
    coordinate separation of $6M$ for models uu1 and sk1--4 and $8M$ for
    models dd1--7. Black hole $1$ is located at $x>0$ and hole $2$ at
    $x<0$. For the spin-kick (sk) runs $\vec{S}_1 = -\vec{S}_2$ and the spins
    are aligned along the $x$-axis, while for all other models $\vec{S}_1 =
    +\vec{S}_2$ and spins are aligned along the $z$-axis. A common horizon
    forms at $t_{\rm cah}$ and the number of gravitational wave cycles from
    $t_0=r_{\rm ex}+50M$ to the peak in the wave amplitude is $N_{\rm
      cyc}$. $E_{\rm EMOP}$ is the total ringdown energy radiated in $l=m=2$
    and $(l=2,m=-2)$; the number in parenthesis is the percentage of energy in
    the $(l=2,m=-2)$ mode.}
\begin{tabular}{l|ccc|cc|cccc}
\hline  \hline
model & $\frac{M_1}{M_2}$ & $\frac{S_1}{M_1^2}$ & $\frac{P}{M}$ &
$\frac{t_{\rm cah}}{M_{\rm ADM}}$ & $N_{\rm cyc}$ &
$10^2\frac{E_{\rm rad}}{M_{\rm ADM}}$ &
$10^2\frac{E_{\rm EMOP}}{M_{\rm ADM}}$ &
$\frac{J_{\rm rad}}{M_{\rm ADM}^2}$ & $\frac{J_{\rm fin}}{M_{\rm fin}^2}$ \\
\hline
uu1 & 1.0 & $0.926$ & $0.126$  & 269.8 & 7.68 &$8.220\pm0.330$  &2.928 (50)  & $0.4273\pm0.0299$ &  0.9505  \\
dd1 & 4.0 & $-0.75$ & $0.0774$ & 129.7 & 2.23 &$0.705\pm0.028$  &0.338 (50)  & $0.0567\pm0.0040$ &  0.0533  \\
dd2 & 4.0 & $-0.80$ & $0.0778$ & 122.5 & 2.09 &$0.677\pm0.027$  &0.362 (50)  & $0.0546\pm0.0038$ &  0.0237  \\
dd3 & 4.0 & $-0.82$ & $0.0779$ & 124.0 & 2.07 &$0.664\pm0.027$  &0.354 (50)  & $0.0539\pm0.0038$ &  0.0122  \\
dd4 & 4.0 & $-0.83$ & $0.0780$ & 121.4 & 2.02 &$0.654\pm0.026$  &0.342 (50)  & $0.0529\pm0.0037$ &  0.0068  \\
dd5 & 4.0 & $-0.84$ & $0.0781$ & 118.3 & 1.95 &$0.646\pm0.026$  &0.326 (50)  & $0.0521\pm0.0036$ &  0.0012  \\
dd6 & 4.0 & $-0.85$ & $0.0781$ & 112.8 & 1.86 &$0.630\pm0.025$  &0.302 (50)  & $0.0504\pm0.0035$ & -0.0038  \\
dd7 & 4.0 & $-0.87$ & $0.0783$ & 111.8 & 1.82 &$0.626\pm0.025$  &0.287 (50)  & $0.0484\pm0.0034$ & -0.0154  \\
\hline
sk1 & 1.0 & $0.727$ & $0.140$  & 109.5 & 2.52 &$3.453\pm0.138$  &2.028 (67)  & $0.2030\pm0.0142$ &  0.6821   \\
sk2 & 2.0 & $0.727$ & $0.125$  & 129.0 & 2.93 &$2.665\pm0.107$  &1.159 (71)  & $0.1591\pm0.0111$ &  -        \\
sk4 & 4.0 & $0.727$ & $0.0903$ & 119.0 & 2.53 &$1.405\pm0.056$  &0.381 (73)  & $0.0727\pm0.0051$ &  -        \\
\hline \hline
\end{tabular}
\end{table}

In Table \ref{tab: models} we list the initial physical parameters of all
binaries of sequence 2. We also list: the time $t_{\rm cah}$ of formation of a
common apparent horizon; the number of wave cycles $N_{\rm cyc}$ derived from
the phase\footnote{To remove the initial radiation burst the phase was
  integrated from $t=50M+r_{\rm ex}$, where $r_{\rm ex}$ is the extraction
  radius, up to the maximum in the wave amplitude.} of the $(l=2,m=2)$ mode;
the total radiated energy $E_{\rm rad}$ and the total radiated angular
momentum $J_{\rm rad}$, excluding again the initial data burst; the energy
radiated in ringdown $E_{\rm EMOP}$, as estimated using the energy-maximized
orthogonal projection \cite{Berti:2007fi}, for the dominant modes $(l=2,m=2$)
and when the symmetry is broken (i.e., for the sk runs) also for $(l=2,m=-2)$;
and the final black hole spin $j_{\rm fin}=J_{\rm fin}/M^2_{\rm fin}$. For
models sk2 and sk4, some angular momentum is radiated in the $x$- and
$y$-directions.  This results in a realignment of the final spin, and
computing $j_{\rm fin}$ becomes more difficult. For this reason, the
corresponding entries in the table are empty.  Following the convention of
Ref.~\cite{Sperhake:2007gu}, initial data are normalized to $M=M_1+M_2$, while
all radiated quantities are normalized to the ADM mass $M_{\rm ADM}$. The
corresponding details for the non-spinning models can be found in
Refs.~\cite{Berti:2007fi, Gonzalez2007}.

In the notation of Sec.~II~E of Ref.~\cite{Sperhake2006}, the grid setup is
$\{(256,128,64,32,16,8)\times(2,1,0.5),~h=1/80\}$ for the uu1 run,
$\{(384,192,96,56,24,12)\times(3,1.5,0.75),~h=1/48\}$ for the sk1 and sk2
runs, and $\{(384,192,96,56,24)\times(6,3,1.5,0.75),~h=1/48\}$ for all other
simulations.  In addition, we have performed higher-resolution simulations
with $h=1/52$ and $h=1/56$ of the two models labelled sk1 and sk2 in Table
\ref{tab: models}.

\begin{figure}[ht]
\centering
\includegraphics[height=7.0cm,angle=-90]{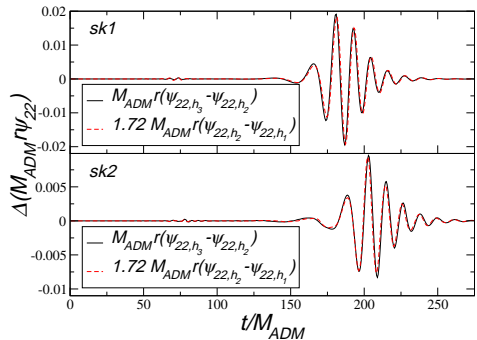}
\caption{Convergence analysis of the $(l=2,m=2)$ mode of $M_{\rm
    ADM}r\Psi_4$ for simulations sk1 and sk2 of Table \ref{tab:
    models}.  The difference between the high-resolution runs has been
  rescaled by 1.72, as expected for sixth-order convergence.  }
\label{fig:convergence}
\end{figure}

The resulting convergence plot for the $(l=2,m=2)$ mode $\Psi_4$ of
the Newman-Penrose scalar $\Psi_4$ extracted at $r_{\rm ex}=60M_{\rm
  ADM}$ is shown in Fig.~\ref{fig:convergence}. The differences
between the high-resolution runs have been rescaled by a factor of
$1.72$, as expected for sixth-order convergence.

To estimate uncertainties arising from finite resolution, we use the
differences obtained for these convergence runs. In order to allow for
the fact that the {\sc Lean} code also contains ingredients of lower
than sixth-order accuracy, we follow the approach of
Ref.~\cite{Berti:2007fi} and apply a Richardson extrapolation assuming
second-order accuracy. Using these conservative estimates, and the
observed
$1/r_{\rm ex}$ fall-off of the errors on radiated energy and momenta,
we obtain very similar uncertainty estimates for simulations sk1 and
sk2. At resolution $h=1/48$ and extraction radius $r_{\rm ex}=60M_{\rm
  ADM}$ these uncertainties are $\sim 4\%$ for the radiated energy and
$\sim 7\%$ for the radiated angular momenta.
%
%
For the discussion below, it is important to remark that our numerical results
{\it underestimate} the radiated quantities in all cases.

\section{Energy distribution for non-spinning binaries}
\label{nospin}

The decomposition of gravitational radiation from non-spinning binaries onto
spin-weighted spherical harmonics (SWSHs) $_{-2}Y_{l\,,m}$ can be found in
\cite{Berti:2007fi}, where it was obtained by projecting the 2.5PN
gravitational waveforms\footnote{The SWSH components of $h_+\,,h_{\times}$ are
  related to the corresponding components of $\Psi_4$ by
  $\left(h_+-ih_{\times}\right)_{l\,,m}= -\psi_{l\,,m}/(m\Omega)^2$, with the
  exception of 2.5PN contributions to the $l=m=2$ component. It has recently
  been pointed out that by using the known expressions of the radiative
  multipoles, instead of the waveforms, more information is available and the
  contribution of each multipole can be computed to higher PN order
  \cite{Kidder:2007rt}.} derived in \cite{Blanchet:1996pi,Arun:2004ff}. The
most notable result of Ref.~\cite{Berti:2007fi} is that the leading-order term
contributing to each multipolar component $\psi_{l\,,m}$ of the Weyl scalar
$\Psi_4$ is proportional to the symmetric mass ratio $\eta=q/(1+q)^2$ when $m$
is even, and to $\eta\,\delta M/M=\eta(M_1-M_2)/M$ when $m$ is odd. More
explicitly:

\begin{eqnarray}
\label{meven}
e^{i\tilde{\phi}}\, Mr\, \psi_{l\,, m}&=&
\eta \sum_{n=0}^{5}g_{l\,,m}^{(n)}(\eta)(M\Omega)^{(8+n)/3}\,, \qquad
m~{\rm even}\,;\\
e^{i\tilde{\phi}}\, Mr\, \psi_{l\,, m}&=&
\eta \frac{\delta M}{M}
\sum_{n=1}^{5}k_{l\,,m}^{(n)}(\eta)(M\Omega)^{(8+n)/3}\,, \qquad
m~{\rm odd}\,.\nonumber
\end{eqnarray}
Here $\tilde \phi$ is an orbital phase, including logarithmic corrections in
the orbital frequency $\Omega$. The precise definition of this phase, as well
as the functional form of the coefficients $g_{l\,,m}^{(n)}(\eta)$ and
$k_{l\,,m}^{(n)}(\eta)$, can be found in Appendix A of \cite{Berti:2007fi}.
Each coefficient in the series represents a PN contribution of order $n/2$ to
the leading-order (Newtonian) prediction.

\begin{figure}[ht]
\centering
\includegraphics[height=6.4cm,angle=-90]{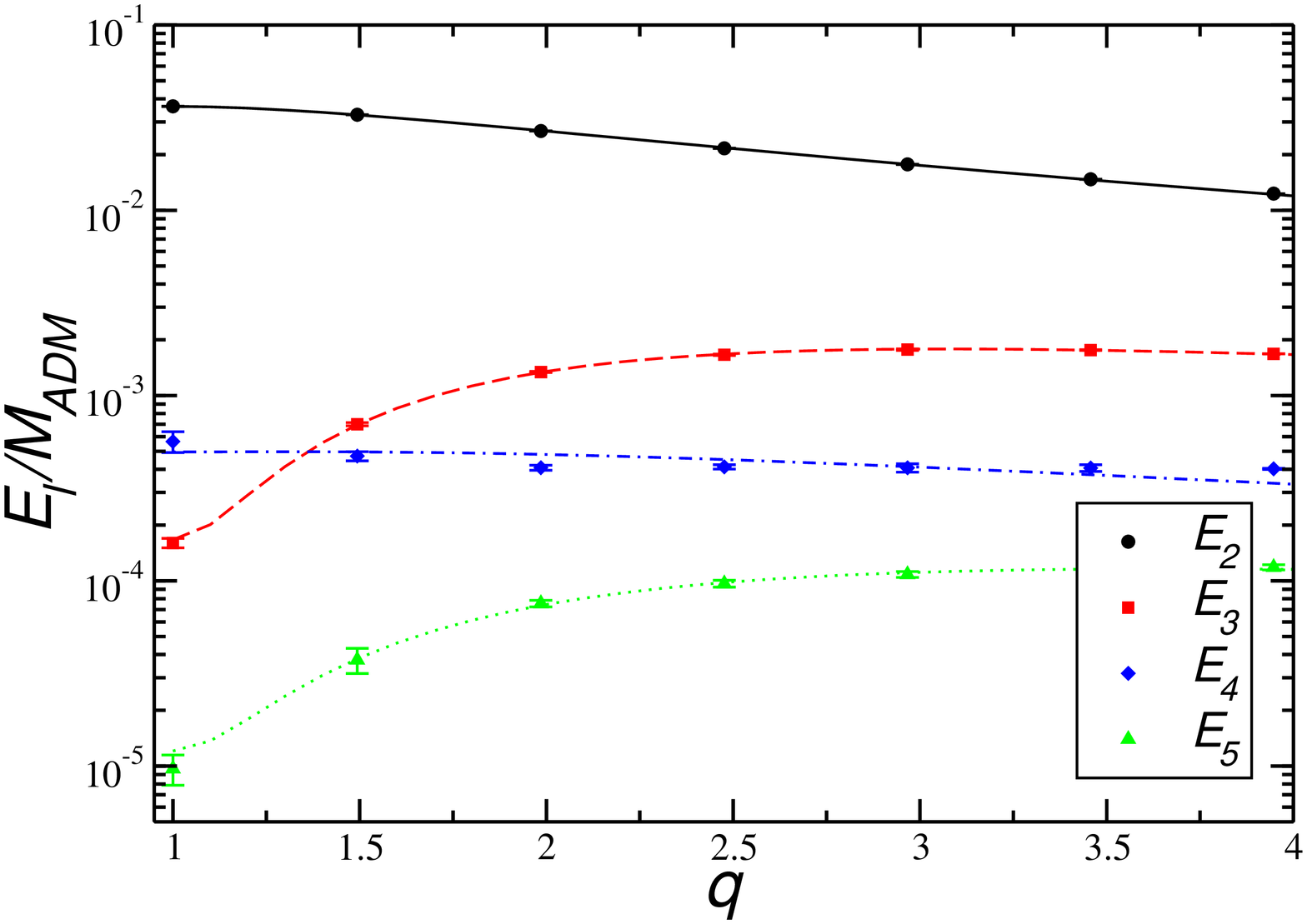}
\includegraphics[height=6.4cm,angle=-90]{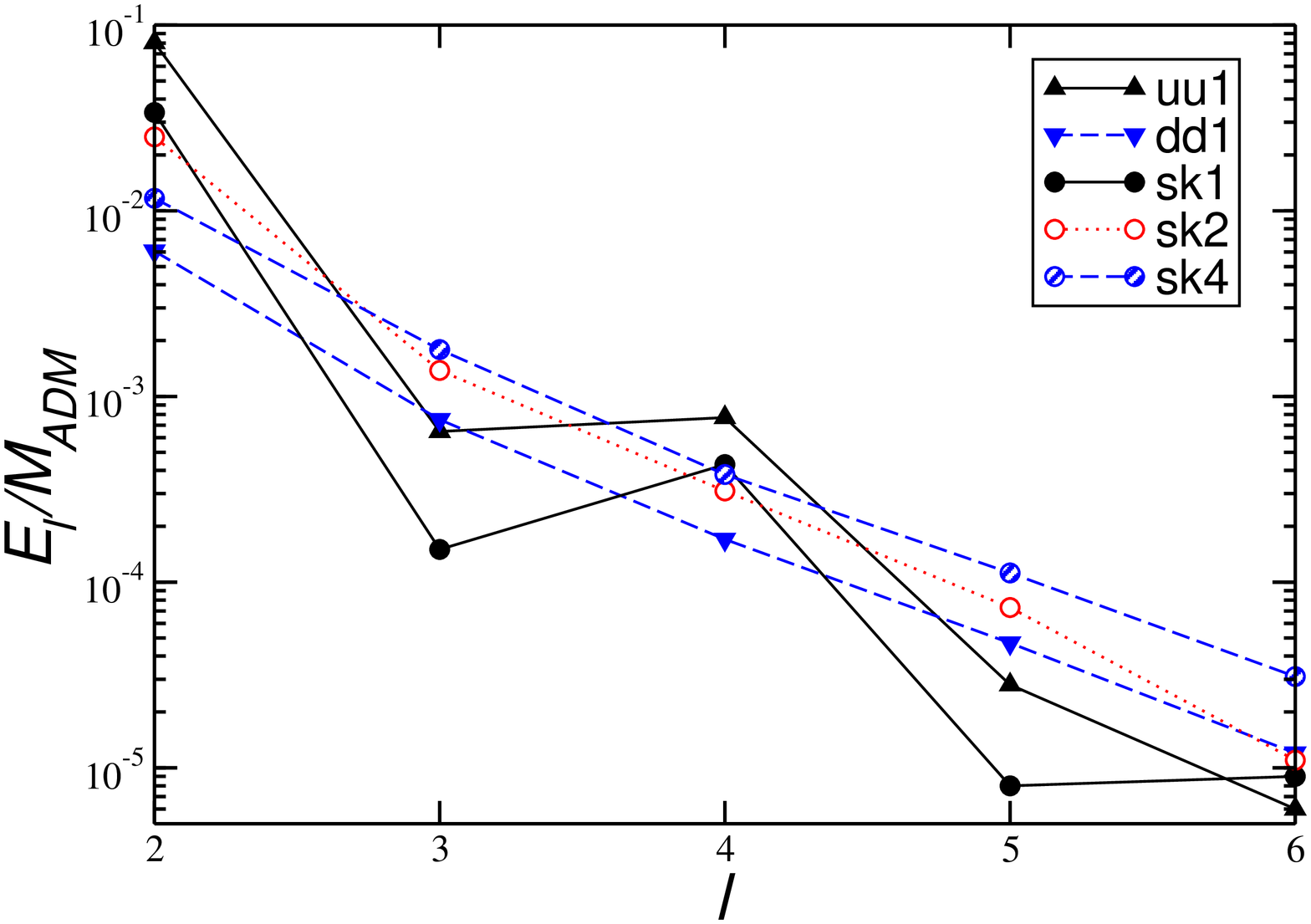}
\caption{Left: Energy $E_l$ in different multipoles for the
  unequal-mass binaries of sequence 1 as a function of $q$ (from
  \cite{Berti:2007fi}).  Right: $E_l$ for some spinning binary
  configurations belonging to sequence 2, as a function of the
  multipole index $l$.  Continuous (black) lines refer to equal-mass
  binaries, the dotted (red) line to a binary with $q=2$, and dashed
  (blue) lines to binaries with $q=4$.}
\label{fig:e}
\end{figure}

From Eq.~(\ref{meven}) it is clear that, for non-spinning binaries,
odd-$m$ multipoles are suppressed in the equal-mass limit.  Since
$l=m$ components typically dominate the radiation for each given $l$,
this also implies that odd-$l$ multipoles are suppressed as $q\to
1$. This is confirmed by numerical simulations of the merger. In
addition, higher multipoles are found to carry a larger fraction of
the total energy as $q$ deviates from unity.  Both of these features
are clear from a glance at the left panel of Fig.~\ref{fig:e}.
In \cite{Berti:2007fi} we also showed that, to leading order, the
total energy radiated in the merger scales like $\eta^2$ and the Kerr
parameter of the final hole scales like $\eta$, providing
phenomenological fits of these quantities.  More general fitting
formulas for the final Kerr parameter, encompassing also binaries with
aligned or anti-aligned spins, can be found in \cite{Rezzolla:2007rd}.

In the right panel of Fig.~\ref{fig:e} we show the energy distribution for
some of our sequence 2 binaries, as a function of the multipole index $l$.
The relative uncertainties of the mode energies have been determined
in analogy to those of the total radiated energy for runs sk1 and sk2
and strongly depend on the energy content (the ``signal strength'').
We find relative uncertainties of about $5\%$ for values of $E_l/M_{\rm ADM}$
down to approximately $5~10^{-4}$. These grow up to
$30\%$ for low signals of $10^{-5}...10^{-4}$. The figure reveals that
even in the presence of spin, odd-$l$ multipoles are suppressed when
$q=1$. As first observed in \cite{Campanelli:2006uy}, the hang-up (uu)
configuration stays in orbit for a longer time and radiates more energy before
merging. On the contrary, our spin-flip (dd) simulations with $q=4$ merge very
rapidly (compare the number of cycles $N_{\rm cyc}$ in Table \ref{tab:
  models}).  All dd simulations radiate roughly the same amount of energy, so
we only show run dd1 in the plot. By comparing the sk1, sk2 and sk4 runs we
confirm that our conclusion in \cite{Berti:2007fi}, that large-$q$ binaries
radiate more energy in higher multipoles, holds true also for these spinning
binaries. This is clear from the slopes of $E_l$ as a function of $l$ for the
different sk runs, and it is nicely illustrated by a comparison of the sk2 and
sk4 runs.  Even though sk4 radiates roughly half the energy radiated by sk2,
the energy radiated in each $l>2$ multipole by the sk4 run is larger.

\section{Leading-order spin-orbit and spin-spin contributions}
\label{spin}

For spins aligned or anti-aligned with the orbital angular momentum,
the leading-order spin contributions to the various multipolar
components are most easily derived by projecting Eqs.~(F24) and
(F25) of Ref.~\cite{Will:1996zj} onto SWSHs, according to the
procedure described in \cite{Berti:2007fi}.  Let $S_i$ be the
projection of the spin of body $i$ on the axis orthogonal to the
orbital plane. $S_i$ is positive (negative) if the spins are aligned
(anti-aligned) with the orbital angular momentum. Define the
dimensionless spin parameter $j_i=S_i/M_i^2$ ($i=1,~2$) and the spin
combinations $\chi_s=(j_1+j_2)/2$, $\chi_a=(j_1-j_2)/2$. Including
only the dominant spin-orbit (1.5PN order) and spin-spin (2PN)
terms, in addition to the non-spinning part of Eq.~(\ref{meven}) we
find three spin-dependent multipolar contributions:
\begin{eqnarray}
Mr\, \psi_{2\,, 2}^{\rm spin}e^{i\tilde{\phi}}&=&
32\sqrt{\frac{\pi}{5}}\eta (M\Omega)^{8/3}\\
\label{c22e} &\times&\left[\frac{4}{3}\left (\chi_s(\eta-1)-\chi_a
\frac{\delta M}{M}\right ) M\Omega +2\eta(\chi_s^2-\chi_a^2)
(M\Omega)^{4/3}\right ]\,,\nonumber\\
Mr\, \psi_{2\,, 1}^{\rm spin}e^{i\tilde{\phi}}&=&
-\frac{8}{3}\sqrt{\frac{\pi}{5}}\eta
(M\Omega)^{3} \left[\frac{3}{2}\left (\chi_s\frac{\delta
M}{M}+\chi_a\right )(M\Omega)^{1/3}\right]\,.
\label{c21e}\\
Mr\, \psi_{3\,, 2}^{\rm spin}e^{i\tilde{\phi}}&=&
\frac{32}{3}\sqrt{\frac{\pi}{7}}\eta
(M\Omega)^{10/3}\left[4\chi_s\eta(M\Omega)^{1/3}\right]\,.
\label{c32e}
\end{eqnarray}

These equations demonstrate that odd-$m$ multipoles do not always
vanish for equal-mass systems.  For example, $\psi_{2\,, 1}^{\rm
  spin}$ contains a term which is {\it not} proportional to the mass
difference $\delta M/M$: for equal masses and $j_1\neq j_2$, the {\em
  dominant} contribution to $\psi_{2\,, 1}$ comes from spin
terms. Therefore, by simulating binaries with equal masses and unequal
spins and by looking at the $(l=2,m=1)$ component we can disentangle
subleading spin effects from the leading-order non-spin contributions,
and thus facilitate the comparison of spin definitions in PN theory
and in numerical simulations.  Unfortunately, for systems with equal
mass {\it and equal spins}, such as our uu1 model, $\psi_{2\,, 1}^{\rm
  spin}=0$.  However, we can still study the effect of spin terms on
the convergence of the PN approximation by considering the $(l=2,m=2)$
component.

To see how this is possible, consider first the non-spinning case. By
taking the modulus of (\ref{meven}) we get a PN series relating the
orbital frequency and the gravitational wave amplitude
$|\psi_{l\,,m}|$.  The convergence rate of the series to the numerical
results can be studied as follows.

First, we observe that the frequency of the gravitational waves $\omega_{\rm
  GW}$ in a multipolar component $\psi_{l\,,m}$ is related to the orbital
frequency $\Omega$ by $\omega_{\rm GW}=m\Omega$.  Therefore, given a
time-dependent component $\psi_{l\,,m}$ of the Weyl scalar, the numerical
value of the binary's orbital frequency can be estimated as $\Omega\simeq
\omega_{Dm}=-{\rm Im} (\dot \psi_{l\,,m}/\psi_{l\,,m})/m$
\cite{Buonanno2006,Berti:2007fi}.  Consider now the modulus of
Eq.~(\ref{meven}), possibly with the addition of spin terms such as
Eq.~(\ref{c22e}). Given $|\psi_{l\,,m}(t)|$ on the left-hand side (as obtained
from the simulation), at any give PN order we can (at least in principle)
numerically invert the PN expansion on the right-hand side. Since this
expansion is only valid for quasi-circular binaries, in this way we get a
post-Newtonian quasi-circular (PNQC) estimate of the orbital frequency:
$\Omega\simeq \omega_{\rm PNQC}$.  If the PNQC approximation works well,
$\omega_{\rm PNQC}$ should be close to $\omega_{Dm}$. Furthermore, if the PN
approximation is converging, the agreement should get better as we increase
the PN order, i.e.  the number of terms in the sum on the right-hand side of
Eq.~(\ref{meven}).

\begin{figure}[ht]
\centering
\includegraphics[height=6.4cm,angle=-90]{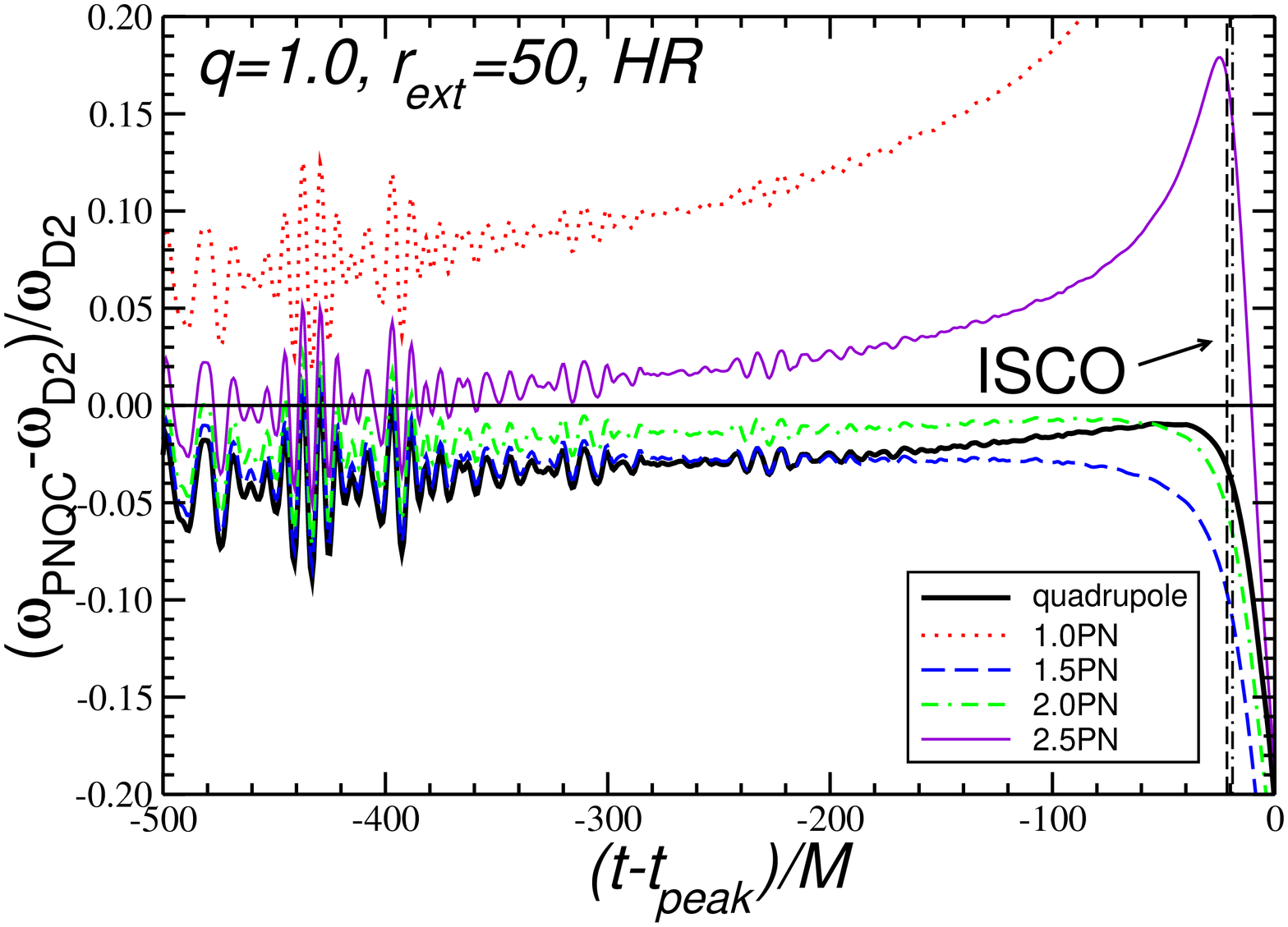}
\includegraphics[height=6.4cm,angle=-90]{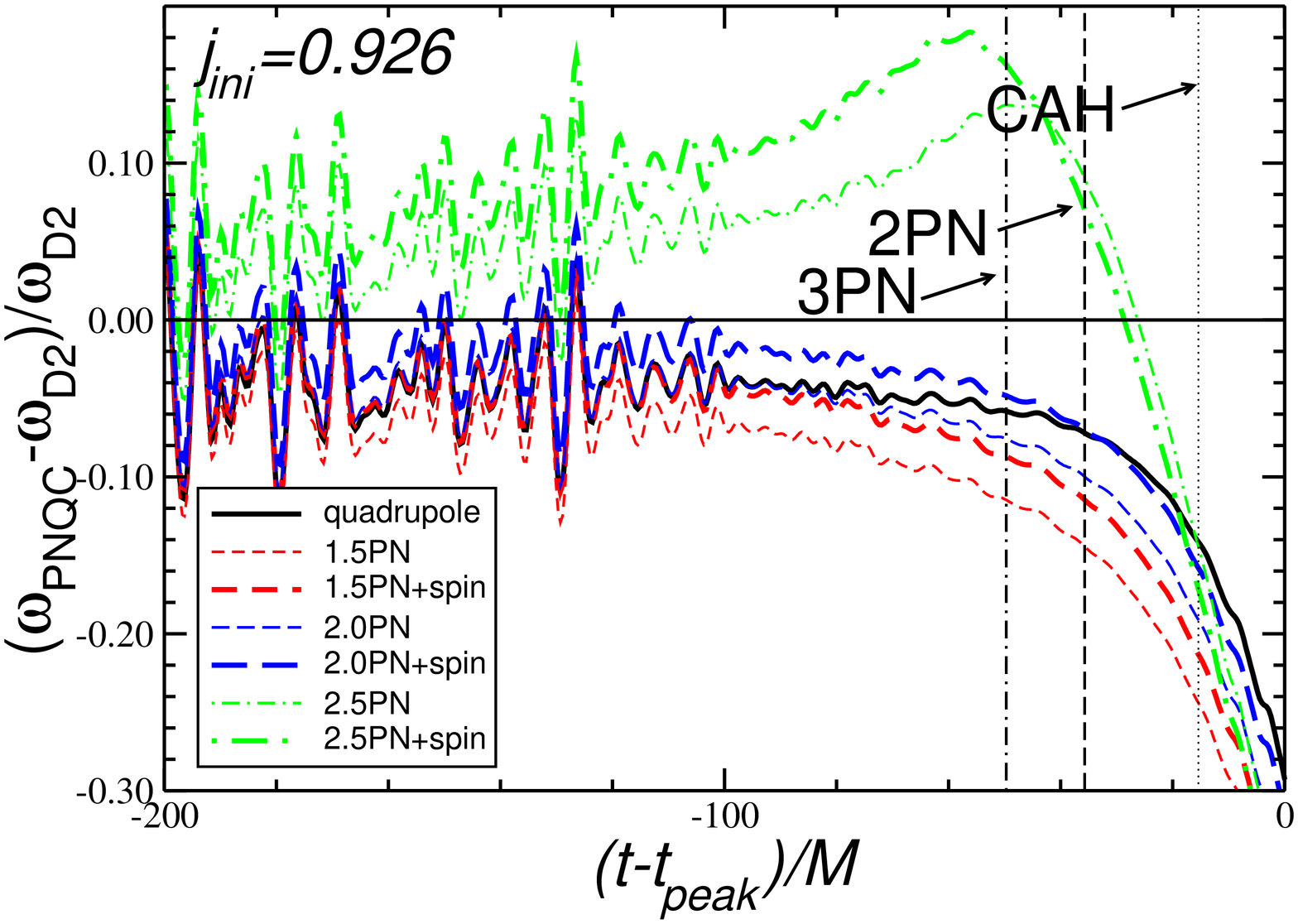}     
\caption{Convergence of the PNQC expansion. Left: non-spinning
  binaries; right: spinning binaries. In the right panel, thick lines
  estimate the PNQC frequency including the spin terms of
  Eq.~(\ref{c22e}), and thin lines omit the spin terms.}
\label{fig:pnqc}
\end{figure}

In Fig.~\ref{fig:pnqc} we show the relative deviation between
$\omega_{\rm PNQC}$ and $\omega_{D2}$, considering the dominant
($l=m=2$) component of the radiation. Let us first consider the left
panel. There we show the relative deviation between $\omega_{\rm
  PNQC}$ and $\omega_{D2}$ for the longest non-spinning, equal-mass
simulation considered in \cite{Berti:2007fi}, at different PN
orders. At early-times we see oscillations in the relative deviation,
that damp away as the binary evolves. The magnitude of the relative
deviation $|(\omega_{\rm PNQC}-\omega_{D2})/\omega_{D2}|$ can be taken
as an indicator of the accuracy of the PN approximation.  The plot
shows that the convergence of the PN series is not monotonic. The
transition from inspiral to plunge is roughly marked by the vertical
lines, that correspond to the point where the orbital frequency equals
the innermost stable circular orbit or ISCO (as defined in
\cite{Buonanno:2002fy}, computed at 2PN and 3PN to bracket
uncertainties). At this point the PNQC frequency, which only makes
sense in the inspiral phase, decouples from $\omega_{D2}$.

The right panel shows the relative deviation between $\omega_{\rm PNQC}$ and
$\omega_{D2}$ for our uu1 run. Vertical lines mark again the 2PN and 3PN ISCO,
that for such large aligned spins corresponds to much higher orbital
frequencies: in our case $M\Omega_{\rm ISCO}^{\rm 3PN, spin}=0.247$, while for
zero spins we would get $M\Omega_{\rm ISCO}^{\rm 3PN,no~spin}=0.129$.
Here we also indicate the formation of a common apparent horizon (CAH).
The fact
that the PNQC estimate again deviates from $\omega_{D2}$ at the ISCO seems to
indicate that the PN notion of orbital instability makes sense even for such
large values of the spin. The relatively short duration of the simulation, and
the large wiggles induced by numerical noise, clearly illustrate the need to
start simulations of spinning binaries at larger separation. It is also clear
that including spin-terms improves the agreement between the numerics and the
PNQC approximation at 1.5PN and 2PN orders. The trend is reversed at 2.5PN,
possibly because we are not including 2.5PN spin contributions. Higher-order
calculations of spin contributions in PN theory and longer simulations will be
necessary for more accurate comparisons.

\section{Producing a Schwarzschild remnant}
\label{Schw}

A question of particular astrophysical importance concerns the final spin
resulting from the inspiral and merger of black-hole binaries with arbitrary
initial parameters \cite{Buonanno:2007sv, Rezzolla:2007rd}. An intriguing
special case is that where two black holes with initial spins anti-aligned
with the orbital angular momentum merge forming a non-spinning hole.

Our sequence of runs with anti-aligned initial spins and mass ratio $q=4$
(symmetric mass ratio $\eta=0.16$) has been designed to bracket the critical
point of formation of a Schwarzschild black hole, as predicted in
Refs.~\cite{Buonanno:2007sv, Rezzolla:2007rd}. We calculate the final Kerr
parameter using energy and angular momentum balance arguments, i.e. we compute
the final black-hole mass as $M_{\rm fin}=M_{\rm ADM}-E_{\rm rad}$ and the
final angular momentum as $J_{\rm fin}=J_{\rm ini}-J_{\rm rad}$. The resulting
dimensionless Kerr parameter $j_{\rm fin}$ is given in the rightmost column of
Table \ref{tab: models}. Applying a linear regression analysis to these
results and the associated error estimates leads to the fitting formula
$j_{\rm fin}=(-0.570\pm0.040)[j_i-(0.842\pm0.003)]$.  A Schwarzschild remnant
is thus produced when the initial spin has the ``critical'' value $j_{\rm
  ini}\simeq -0.842\pm0.003$.  As mentioned above, we generally find the
uncertainties due to finite differencing and extraction radius to
underestimate the radiated angular momentum, so that we expect the correct
value to be $>-0.842$.
We can compare our result to that of Ref.~\cite{Rezzolla:2007rd} by applying
standard error propagation to the uncertainties listed in their Eq.~(7).
Specifically, we solve their Eq.~(6) for $a$, set $a_{\rm fin}=0$ and
calculate the uncertainty $\Delta a^2 =\sum_i (\partial a/\partial v_i)^2
\Delta v_i^2$, where $v_i = s_4,~s_5,~t_0,~t_2$ and $t_3$. The resulting
critical angular momentum of $0.824\pm 0.019$ agrees very well with our value.
Both results also agree well with that of $-0.815$ predicted in
Ref.~\cite{Buonanno:2007sv}.

Finally, a systematic uncertainty in our numerical results is due to the
relatively small initial separation of the holes. However, a comparison of
non-spinning simulations starting at different initial separations shows that
most of the angular momentum is radiated during the last orbit prior to
formation of a common apparent horizon \cite{Berti:2007fi}. Therefore this
systematic error should not significantly affect our results. Simulations
starting at larger separation are required to verify this expectation, and we
plan to investigate this effect in future work.

\section{Ringdown energy}
\label{EMOP}

Present ringdown searches in LIGO data are based on matched filtering.  For
this reason, a practical criterion to define the ringdown content of a given
waveform is the energy-maximized orthogonal projection (EMOP) discussed in
\cite{Berti:2007fi}, which is basically matched filtering in white noise.  As
shown in \cite{Berti:2007fi}, the EMOP estimate of the energy radiated in
ringdown is a lower bound on the energy that can be detected by matched
filtering. Table \ref{tab: models} lists the sum of the ringdown energies
radiated in the dominant $(l=2,m=2)$ and $(l=2,m=-2)$ components of the
radiation. For the sk runs the radiation is not symmetric with respect to the
$z$-axis \cite{Brugmann:2007zj}, therefore we also list (in parentheses) the
percentage of $E_{\rm EMOP}$ radiated in $(l=2,m=-2)$.  From the data we see
that as much as $\sim 3\%$ of the rest energy of the system is radiated in the
ringdown phase for equal-mass binaries. From our previous study of
non-spinning binaries \cite{Berti:2007fi} it is reasonable to assume that, for
fixed initial Kerr parameters, $E_{\rm EMOP}\sim \eta^2$. Such high ringdown
efficiencies are good news for the detection of ringdown waves and for their
use in parameter estimation \cite{Berti:2005ys, Berti:2007zu}.

EMOP estimates for runs sk2 and sk4 should be taken with caution. For these
runs, the spin axis of the final black hole is tilted with respect to the
$z$-axis of the coordinate frame used for the evolutions and for wave
extraction, and our chosen reference frame is not appropriate to describe the
symmetries of the final hole.  The tilt in the final spin angle produces mode
mixing in the ringdown phase: a pure $(l,m)$ mode in the frame chosen for
numerical computations is a sum of modes with different $(l',m')$'s in the
frame whose $z$-axis coincides with the symmetry axis of the final hole, and
vice versa. Because of this mode mixing, an estimate of the Kerr parameter of
the final hole using quasinormal mode fits is not trivial. A detailed
treatment of this problem will be presented in future work.

\section*{Acknowledgments}
We thank Clifford Will for useful correspondence and for pointing out some
typos in the Appendix of Ref.~\cite{Will:1996zj}. This work was supported in
part by DFG grant SFB/Transregio~7 ``Gravitational Wave Astronomy''. E.B.'s
research was supported by an appointment to the NASA Postdoctoral Program at
the Jet Propulsion Laboratory, California Institute of Technology,
administered by Oak Ridge Associated Universities through a contract with
NASA; by the National Science Foundation, under grant number PHY 03-53180; and
by NASA, under grant number NNG06GI60 to Washington University.  V.C.'s work
was partially funded by Funda\c c\~ao para a Ci\^encia e Tecnologia (FCT) -
Portugal through projects PTDC/FIS/64175/2006 and POCI/FP/81915/2007.  We
thank the DEISA Consortium (co-funded by the EU, FP6 project 508830).
Computations were performed at LRZ Munich.  We are grateful to CFC in Coimbra
for granting us access to the Milipeia cluster.


\section*{References}


\end{document}